\documentstyle[12pt]{article}
\input epsf

\newcommand{\balpha}{{\mbox{\boldmath $\alpha$}}}

\newcommand{\bbeta}{{\mbox{\boldmath $\beta$}}}

\newcommand{\bphi}{{\mbox{\boldmath $\phi$}}}

\newcommand{\be}{\begin{equation}}
\newcommand{\ee}{\end{equation}}
\newcommand{\bea}{\begin{eqnarray}}
\newcommand{\eea}{\end{eqnarray}}
\newcommand{\ba}{\begin{array}}

\newcommand{\ea}{\end{array}}

\def\bbox{{\,\lower0.9pt\vbox{\hrule \hbox{\vrule height 0.2 cm
\hskip 0.2 cm \vrule height 0.2 cm}\hrule}\,}}
\newcommand{\dsl}{\pa \kern-0.5em /}

\newcommand{\ep}{\epsilon}

\newcommand{\nn}{\nonumber \\}

\font\mybb=msbm10 at 10pt
\def\bb#1{\hbox{\mybb#1}}
\def\bZ {\bb{Z}}

\begin{document}

%%%%%%%%%%%%%%%% title page %%%%%%%%%%%%%%%%%%%%%%%%%%%%%%%%%%%%

\begin{titlepage}
\vfill
\begin{flushright}
KIAS-P99103\\
QMW-PH-99-15\\
IASSNS-HEP-99/110\\
hep-th/9912082\\
\end{flushright}

\vfill

\begin{center}
\baselineskip=16pt
{\Large\bf Monopole Dynamics and BPS Dyons} \\
{\Large\bf in $N=2$ Super-Yang-Mills Theories}
\vskip 10.mm
{ ~Jerome P. Gauntlett$^{*,1}$, ~{}Nakwoo Kim$^{*,2}$,
~{}Jaemo Park$^{+,3}$, and ~Piljin Yi$^{\#,4}$ } \\
%\\[2mm]
\vskip 0.5cm
%\vfill
{\small\it
$^*$
  Department of Physics, Queen Mary and Westfield College\\
  Mile End Rd, London E1 4NS, UK
}\\
\vspace{6pt}
{\small\it
$^+$
  School of Natural Sciences, Institute for Advanced Study\\
  Princeton, NJ 08540, USA
}\\
\vspace{6pt}
{\small\it
 $^\#$
School of Physics, Korea Institute for Advanced Study\\
207-43, Cheongryangri-Dong, Dongdaemun-Gu, Seoul 130-012, Korea
}
\end{center}
\vfill
\par
\begin{center}
{\bf ABSTRACT}
\end{center}
\begin{quote}
We determine the low energy dynamics of monopoles in pure $N=2$ Yang-Mills 
theories for points in the vacuum moduli space where the two Higgs fields 
are not aligned. The dynamics is governed by a supersymmetric quantum mechanics
with potential terms and four real supercharges. The corresponding superalgebra
contains a central charge but nevertheless supersymmetric states preserve all 
four supercharges. The central charge depends on the sign of the electric 
charges and consequently so does the BPS spectrum. We focus on the $SU(3)$ 
case where certain BPS states are realised as zero-modes of a Dirac 
operator on Taub-NUT space twisted by the tri-holomorphic Killing vector field.
We show that the BPS spectrum includes hypermultiplets that are consistent 
with the strong- and weak-coupling behaviour of the Seiberg-Witten theory.

\vfill
\vskip 5mm
\hrule width 5.cm
\vskip 5mm
{\small
\noindent $^1$ E-mail: j.p.gauntlett@qmw.ac.uk \\
\noindent $^2$ E-mail: n.kim@qmw.ac.uk\\
\noindent $^3$ E-mail: jaemo@sns.ias.edu \\
\noindent $^4$ E-mail: piljin@kias.re.kr\\
}
\end{quote}
\end{titlepage}
%%%%%%%%%%%%%%%%%%%%%%%%%%%%%%%%%%%%%%%%
\setcounter{equation}{0}

\section{Introduction}

The BPS spectrum of monopoles and dyons are an important non-perturbative 
feature of supersymmetric Yang-Mills theories. At weak coupling
one can determine the BPS spectrum using semi-classical techniques.
Following \cite{gaunt,blum},
the BPS spectrum of $N=2$ and $N=4$ theories was studied 
in a number of papers \cite{sen,sethi,lwy,gibbons}
at points in the moduli space of vacua where only a single Higgs field 
was involved, or more precisely where all of the Higgs fields were aligned.
In these cases one studies certain supersymmetric quantum mechanics models
with the target manifold given by the moduli space of classical BPS
monopole solutions.

New features arise when one studies the spectrum at points in the moduli
space where the Higgs fields are not aligned
\cite{hf,bergman,hashimoto,leeyi,bak,tong}. 
For theories with $N=4$ supersymmetry, the BPS bound is 
determined by two complex central charges 
that appear in the supersymmetry algebra. For aligned Higgs fields
the two charges are necessarily equal and a BPS state preserves 1/2 of
the supersymmetry. When the six Higgs fields are not aligned the central
charges can be different and then the BPS states preserve 1/4 of 
the supersymmetry.

The low energy dynamics of monopoles for non-aligned Higgs fields
in $N=4$ theories were recently studied by Bak. et. al \cite{blly}. 
The supersymmetric quantum mechanics is still based 
on the same BPS monopole moduli space, but is now supplemented
by a supersymmetric potential term which is constructed from a set of
tri-holomorphic Killing vector fields that generate unbroken $U(1)$ gauge
symmetries. It was noticed in Ref.~\cite{tong} that this potential
naturally appears in the expression for the energy of BPS states,
while Bak et.al. later
showed how the same potential occurs in the low energy dynamics,
albeit with an important multiplicative factor of $1/2$, and used the
resulting Lagrangian to study the spectrum of 
$N=4$ Yang-Mills, including 1/4 BPS states.

In this paper we will analyse analogous issues for pure $N=2$ supersymmetric
Yang-Mills theories.
Since the $N=2$ supersymmetry algebra has
one complex central charge there can only be BPS states preserving 
1/2 of the supersymmetry. 
Since the pure $N=2$ Yang-Mills theory can be embedded
in the $N=4$ theory it is not surprising that the central charge is one of
the central charges that appear in the $N=4$ theory. It is interesting
that the other $N=4$ central charge also appears as a bound on the classical
mass of dyons, but it is no longer related to preservation of
supersymmetry. If this latter bound is stronger than the BPS bound for
a given set of charges, then no BPS state can exist with those charges.

At points in the vacuum moduli space of pure $N=2$ Yang-Mills theories
where the Higgs fields are aligned, the low energy dynamics is a
supersymmetric quantum mechanics on the moduli space of BPS monopoles
with four real supersymmetries \cite{gaunt}. The BPS
states correspond to harmonic spinors on the monopole moduli space
(or equivalently, on a hyperK\"ahler manifold,
harmonic holomorphic forms). This follows from the simple fact that one of
the low energy supercharges $Q$ is proportional to a Dirac operator
on the moduli space,
\begin{equation}
{\cal D}=-i\gamma^\mu\nabla_\mu ,
\end{equation}
with covariant derivative on the moduli space $\nabla$,
and its square gives the supersymmetric
sigma-model Hamiltonian.
\begin{equation}
Q^2=\frac{1}{2}\,{\cal D}^2={\cal H}_0 .
\end{equation}

With two Higgs fields active, we will argue that the low energy dynamics  
includes a supersymmetric potential term as in the
$N=4$ theories \cite{blly}. 
The only difference
from the $N=4$ case is that the number of fermions and the number of
supercharges is reduced by half. We will write down this low energy
Lagrangian explicitly in section 3. The supercharge is now given by
the Dirac operator twisted by a tri-holomorphic vector field $G$. The
superalgebra then has the general form,
\begin{equation}
Q^2=\frac{1}{2}\,({{\cal D}-\gamma^\mu G_\mu })^2={\cal H}-{\cal Z} ,
\end{equation}
where ${\cal H}$ is the modified Hamiltonian,
and ${\cal Z}$ is a real central charge defined by the Lie
derivative along $G$,
\begin{equation}
{\cal Z}=-i{\cal L}_{G} ,
\end{equation}
and measures a linear combination of electric charges.

The BPS states with ${\cal H}={\cal Z}$ 
preserve not only the supercharge $Q$ but, as we will
show, all four supercharges. This is consistent with preservation of 1/2 of 
the spacetime supersymmetry. It is interesting to note that if we flip
the signs of the 
electric charges so that ${\cal Z}\to -{\cal Z}$, the state will
no longer be BPS. This should be 
contrasted with the $N=4$ theory, where BPS states
with electric charges of both signs may occur and break a further half of
the supersymmetries in general.

We will analyze in some detail the simplest case of $SU(3)$ broken 
to $U(1)\times U(1)$ by two adjoint Higgs. In particular we will focus on
BPS states with a $(1,1)$ magnetic charge. The
BPS monopole moduli space for this case is given by $R^3\times(R\times M)/Z$ 
where $M$ is Taub-NUT space \cite{lwy,connell}. The BPS spectrum
is then determined by solving the Dirac equation on the Taub-NUT manifold 
twisted by the tri-holomorphic Killing vector field 
and we will be able to utilise the 
results of Pope who studied precisely the same operator in 
\cite{pope}.

An early analysis of the BPS spectrum of $N=2$ $SU(3)$ Yang-Mills theories in 
the weak coupling regime was carried out in the context of Seiberg-Witten 
theories \cite{sw} by Fraser and
Hollowood \cite{hollo}. Acting with semi-classical monodromy transformations 
on purely magnetic states, they found new states 
in a certain part of the vacuum moduli space
whose magnetic charge is $(1,1)$ 
and whose electric charge is $(n,n-1)$ with arbitrary integer $n$. Since
the monodromy cannot alter the supermultiplet structures, all of these dyons
fill out hypermultiplets. By solving
the low energy dynamics of two distinct monopoles, we will find that
these are a particular case of more general states with electric charges,
$(m,l)$ where integers $m$ and $l$ are such that $m> l$. The size of
supermultiplet of the BPS state grows linearly with the 
positive integer $m-l$. 

This paper is organised as follows. Section 2 will briefly summarise
the classical energy bound of the pure $N=2$ Yang-Mills theory. We will
show that there are two bounds on the classical energy and only one of them
corresponds to a BPS bound. In section 3 we will present the 
supersymmetric quantum mechanics with potential that should describe 
the low-energy dynamics of monopoles and dyons.
We analyze the conditions for preserved 
supersymmetry and use this in section 4 to analyze the BPS
spectrum for the case of $SU(3)$. In Section 5, we summarise some of the
previously known results on the spectrum of pure $SU(3)$ Seiberg-Witten 
theory from monodromies as well as strong coupling singularities in the 
vacuum moduli space, and show that the results are
consistent with those in Section 4. We conclude in section 6.

\section{BPS bound}

The $N=2$ super-Yang-Mills Lagrangian is given by
\bea\label{susylag}
L&=&{1\over 2}
tr\Biggl\{-{1\over 2}F_{\mu\nu}F^{\mu\nu}+D_\mu\phi^ID^\mu\phi^I 
+e^2[\phi^1,\phi^2]^2 \nonumber \\ && 
\hskip 1cm + i\bar\chi\Gamma^\mu D_\mu\chi -e\bar \chi[\phi^1,\chi]
-ie\bar\chi\gamma_5[\phi^2,\chi]\Biggr\} ,
\eea
where $\phi^I$, $I=1,2$ denote the two real Higgs fields,
$D_\mu\phi^I=\partial_\mu\phi^I-ie[A_\mu,\phi^I]$ and $\chi$ is a Dirac spinor
and all fields are in the adjoint representation of the gauge group $G$.
The vacuum moduli space demands that $[\phi^1,\phi^2]=0$;
we may choose the asymptotic values of the Higgs fields along
the positive z-axis, say, to be in the Cartan sub-algebra,
$\phi^I={\bphi}^I\cdot {\bf H}$, where ${\bphi}^I$ are vectors of
dimension $r$=rank($G$). This does not completely fix the gauge transformations
as one has the freedom to perform discrete gauge transformations by elements
of the Weyl group. These can be fixed by demanding, 
for example, that
${\bphi^1}\cdot {\bbeta^a}\ge 0$ 
for a given set of simple roots ${\bbeta^a}$ of the Lie algebra $\cal G$ of
$G$. We will only consider points in
the moduli space of vacua where the symmetry is maximally broken to
$U(1)^r$.

For a given vacuum
we can define electric and magnetic charge two-vectors
\be
\vec{Q}_e = tr\oint dS^i E_i \vec{\phi}\,,
\qquad
\vec{Q}_m = tr\oint dS^i B_i \vec{\phi}\, ,
\ee
with $i=1,2,3$ and $\vec{\phi}=(\phi^1,\phi^2)$.
These can be written as
\be
{Q}^I_e = {\bphi}^I\cdot {\bf q}\,,
\qquad
{Q}^I_m ={\bphi}^I\cdot{\bf g}\, ,
\ee
where we have introduced the electric and magnetic
charge vectors given by
\bea
{\bf q}&=&en^a_e {\bbeta^a} ,\nn
{\bf g}&=&{4\pi \over e} n^a_m{\bbeta_a^*} ,
\eea
respectively, where ${\bbeta^a}$ are the simple roots and
${\bbeta_a^*}$ are the simple co-roots of $\cal G$, and
$n^a_m$ are the topological winding numbers and $n^a_e$
are, in the quantum theory, the electric quantum numbers.

By determining the central charges that appear in the supersymmetry 
algebra as in
\cite{wittenolive} we can determine the BPS bound:
\be\label{bpsnequalstwo}
M\ge  |Z_- =(Q^1_e-Q^2_m)+ i(Q^1_m + Q^2_e)| .
\ee
Note that if we introduce a complex rescaled Higgs vector 
${\bf A} =e({\bphi}^1 +i {\bphi}^2)$ and rescale the
charge vectors via 
$\hat {\bf q}={\bf q}/e$ and $\hat {\bf g}=(e/4\pi) {\bf g}$
then the BPS condition becomes 
$M=|{\bf A}\cdot \hat{\bf q} + {\bf A_D}\hat\cdot {\bf g}|$ where
${\bf A_D} =(i4\pi/e^2) {\bf A}$ which is the form
familiar from Seiberg-Witten theory (for vanishing $\theta$) \cite{sw}.

It is illuminating to rederive the BPS bound using Bogomol'nyi's method of
rewriting the energy as a sum of squares plus conserved charges. Indeed we
will see that this gives rise to two bounds on the classical energy.
Since the bosonic part of the $N=2$ Lagrangian differs
from the $N=4$ theory only in
the fact that there are two Higgs fields instead of six Higgs,
one can immediately adapt the derivation of
general BPS bound for the $N=4$ theory
\cite{hf,leeyi} to the $N=2$ case. One finds that
the most stringent bound on the mass is given by
\bea\label{bpse}
M&\ge& 
\sqrt{|\vec{Q}_e|^2 + |\vec{Q}_m|^2 + 2 |\vec{Q}_e||\vec{Q}_m|\sin\xi}\nn
&=&{\rm Max}\;
(\sqrt{|\vec{Q}_e|^2 + |\vec{Q}_m|^2 \pm 2 (Q^2_mQ^1_e-Q^1_mQ^2_e)}) ,
\eea
where $0\le\xi\le\pi$ is the angle between the two 2-vectors $\vec{Q}_e$ and
$\vec{Q}_m$.
This is equivalent to
\be\label{tinky}
M\ge  {\rm Max}\:|Z_\pm =(Q^1_e\pm Q^2_m)+ i(Q^1_m\mp Q^2_e)| .
\ee

In $N=4$ theories, $Z_\pm$ appear as central charges in
the supersymmetry algebra. If a state saturates the 
BPS bound (\ref{tinky}) it will preserve 
1/4 of the supersymmetry. In cases where $Z_+=Z_-$, which occurs when
the angle between $\vec{Q}_e$ and $\vec{Q}_m$ vanishes,
the state will preserve 1/2 of the supersymmetry.
By contrast, in $N=2$ theories there is only
one complex central 
charge that appears in the supersymmetry algebra\footnote{If $\phi^2\to
-\phi^2$ in (\ref{susylag}) the central charge would be
$Z_+$.}, $Z_-$, giving
rise to the BPS bound (\ref{bpsnequalstwo}). A state saturating this bound will
preserve 1/2 of the supersymmetry. A classical soliton can only saturate 
the larger of the two bounds, $|Z_\pm|$. Thus, if it so happens that 
$|Z_-|< |Z_+|$ then there can be no classical BPS soliton with such
charges in such a vacuum. 
In particular, suppose that a state of charge $({\bf g},{\bf q})$ saturates
the BPS bound $|Z_-| > |Z_+|$. Then, for a state of charge 
$({\bf g},-{\bf q})$, the BPS bound $|Z_-|$ will be smaller 
than the classical energy bound $|Z_+|$. 
In the asymptotic region of vacuum moduli space
where a semi-classical analysis is suitable
the quantum corrections to the classical soliton mass will be small
and we conclude that that the latter state cannot be BPS saturated.
This asymmetry with respect to the sign of the electric
charge is a generic feature of the $N=2$ dyon spectrum. This feature 
will be manifest in the low energy superalgebra derived in section 3 and
will be further analysed for the specific example of $SU(3)$ in section 4.

With this knowledge in mind, let us continue exploring the energy bound
further. Defining the linear combinations of Higgs via
\bea
a&=&\cos\alpha\phi^1-\sin\alpha\phi^2 ,\nn
b&=&\sin\alpha \phi^1+\cos\alpha\phi^2 ,
\eea
using the arguments of \cite{hf,leeyi}
the mass bound is saturated when
\bea\label{bpsbound}
E_i&=&\pm D_ia ,\nn
B_i&=&D_ib ,
\eea
and the angle $\alpha$ is constrained to be 
\be\label{anglevalue}
\tan\alpha = {Q_m^1 \mp Q_e^2\over Q_m^2 \pm Q_e^1}\, .
\ee
In addition, in the gauge $A_0=-a$ all fields are static and
Gauss' Law becomes
\be\label{gausslaw}
D^2a-e^2[b,[b,a]]=0 .
\ee
Note that the second equation in (\ref{bpsbound}) is the usual
BPS equation for a single Higgs field.

In terms of the vectors ${\bf a},{\bf b}$, the mass bound can then be written
\be
M\ge {\rm Max}\:(\pm{\bf a}\cdot {\bf q} +{\bf b}\cdot{\bf g} )  ,
\ee
and the constraint (\ref{anglevalue}) is replaced with
\be
{\bf a}\cdot{\bf g} =\pm{\bf b}\cdot{\bf q} .
\ee
It should be emphasised that ${\bphi}^I$ and not ${\bf a}, {\bf b}$
specify the point in vacuum moduli space where a
semi-classical analysis is relevant\footnote{This can be further illustrated
for $N=4$ Yang-Mills. To ensure a duality invariant BPS mass formula,
${\bphi}^I$ are invariant whilst ${\bf a}, {\bf b}$ transform under 
$SL(2,\bZ)$ duality because the angle $\alpha$ transforms.}
since the latter depend on ${\bf g},{\bf q}$ via the angle $\alpha$.

Note that for gauge group $SU(2)$, in order that $[\phi^1,\phi^2]=0$,
$\phi^1$ must be proportional to $\phi^2$.
For finite energy configurations we then
deduce that $\vec{Q}_e$ is proportional to
$\vec{Q}_m$ and hence the only bound on the mass is the
BPS bound given by
$M^2\ge (\vec{Q}_e)^2+(\vec{Q}_m)^2$ as in \cite{wittenolive}. 
It is perhaps worth commenting
that even for gauge group $U(1)$ there are infinite energy configurations
with $\vec{Q}_e$ not proportional to $\vec{Q}_m$ \cite{gkmtz}.

\section{Low-Energy Dynamics of Monopoles and Dyons}

For a single adjoint Higgs field, it is well known that classical 
bosonic dyons can be described as monopoles with some internal 
momentum excited. The low-energy dynamics is determined by a sigma-model whose
classical orbits are geodesics on the moduli space of monopole solutions.
In the case of maximal symmetry breaking, the moduli space
has $U(1)^r$ symmetry, arising from global gauge transformations,
and the corresponding momenta are the conserved
electric charges. 

It was argued in \cite{leeyi} that one can similarly analyze solutions 
of (\ref{bpsbound}),(\ref{gausslaw}) by constructing a modified
low-energy dynamics on monopole moduli spaces. 
Given a solution of the BPS equation $B=Db$, the other BPS equation 
(\ref{gausslaw}) is solved by any gauge-zero mode of the solution.
In the case of widely separated fundamental monopoles \cite{ejw} with
respect to $b$, the solution can be thought of as classically bound
dyons (with respect to $b$). 
Using this information, it was argued that the correct low-energy dynamics is
determined by the sigma-model supplemented with a potential term. 
The arguments presented in
\cite{leeyi} were in the context of $N=4$ theories, but since only
two Higgs fields were involved the arguments can be
immediately adapted to the $N=2$ case.  
In particular, the bosonic Lagrangian should be given by
\begin{equation}
{\cal L}= \frac{1}{2}\, g_{\mu\nu}\dot z^\mu \dot z^\nu
-\frac{1}{2}\,g_{\mu\nu}G^\mu G^\nu ,
\end{equation}
where $g$ is the metric on the monopole moduli space, and  $G$ is a
tri-holomorphic Killing vector field on the moduli space which is
associated with a certain unbroken $U(1)$ gauge symmetry. More precisely,
$G$ is given by
\begin{equation}\label{geedef}
G=e\,{\bf a}\cdot {\bf K} ,
\end{equation}
where the $r$ Killing vectors $K_a$ are generated by $U(1)^r$ unbroken
gauge group acting on the moduli space. 

This low energy dynamics is a nonrelativistic approximation, so
we always assume slow motion in the moduli space of monopoles. A related
but independent condition that is needed to justify the above dynamics
is that the potential energy contribution be small compared to the
rest mass of the monopoles. In particular, when we realise
dyons as bound states of monopoles, the low energy approximation is valid
only if the following condition holds,
\begin{equation}
{\bf a}\cdot{\bf q} \ll {\bf b}\cdot{\bf g} ,
\end{equation}
which is satisfied for weak coupling, since the left hand side $\sim e$
while the right hand side $\sim 1/e$.

The above bosonic dynamics must be generalised to include fermions and
supersymmetry. Monopoles preserve half of the $N=2$ supersymmetry 
in four dimensions,
so the low energy dynamics will have exactly four real supercharges.
In the absence of the potential term (i.e., when only one Higgs is
active), the dynamics has been derived and takes the following form 
\cite{gaunt},
\be
{\cal L}={1\over 2} \biggl( g_{\mu\nu} \dot{z}^\mu \dot{ z}^\nu +
ig_{\mu\nu} \lambda^\mu D_t \lambda^\nu \biggr),
\ee
where $D_t\lambda^\mu=\dot \lambda^\mu
+\Gamma^\mu_{\nu\rho}\dot z^\nu\lambda^\rho$.
The addition of the bosonic potential $G^2/2$ induces a term involving
fermions, and the full supersymmetric Lagrangian with potential is given by
\be
{\cal L}={1\over 2} \biggl( g_{\mu\nu} \dot{z}^\mu \dot{ z}^\nu +
ig_{\mu\nu} \lambda^\mu D_t \lambda^\nu
 - g^{\mu\nu} G_\mu G_\nu - iD_\mu G_\nu  \lambda^\mu \lambda^\nu  \biggr).
\label{action}
\ee
Assuming that the target is
hyperK\"ahler and that the Killing vector field $G$ is tri-holomorphic,
the action is invariant under the following four
supersymmetry transformations
\bea
\delta z^\mu &=& -i\ep\lambda^\mu +i\ep_a {J^{(a)\mu}}_\nu \lambda^\nu ,\nn
\delta \lambda^\mu&=&(\dot z^\mu -G^\mu)\ep +{J^{(a)\mu}}_\nu(\dot z^\nu -
G^\nu)\ep_a
-i\ep_a \lambda^\rho \lambda^\nu {J^{(a)\sigma}}_\rho \Gamma^\mu_{\sigma\nu} ,
\eea
where $\ep,\ep_a$ are constant one component Grassmann odd parameters.
Note that the two-form $dG$ is (1,1) with respect to all complex structures
when $G$ is tri-holomorphic. This in turn implies that $dG$ is anti-self-dual.
The commutator of  two different supersymmetry transformations vanishes,
while those of like supersymmetry transformations gives rise to
a combination of a time translation and the symmetry generated by the
Killing vector field $G$:
\bea
\delta z^\mu &=& k G^\mu ,\nn
\delta \lambda^\mu&=&k {G^\mu}_{,\nu}\lambda^\nu .
\eea
This supersymmetric quantum mechanics thus has all the features we 
require and on this basis we will assume that it is in fact correct.

To quantise we first introduce a frame $e^A_\mu$ and define
$\lambda^A=\lambda^\mu e_\mu^A$ which commute with all bosonic variables.
The remaining canonical commutation relations are then given by
\bea
[z^\mu,p_\nu]&=&i\delta^\mu_\nu  , \nn
\{\lambda^A,\lambda^B\}&=&\delta^{AB} .
\eea
We can realise this algebra on spinors on the moduli space by
letting $\lambda^A=\gamma^A/{\sqrt 2}$, where $\gamma^A$ are gamma matrices.
Since the moduli space is hyperK\"ahler an equivalent quantization
is obtained using holomorphic differential forms.
The supercovariant momentum operator defined by
\be
\pi_\mu=p_\mu-{i\over 4}\omega_{\mu AB}[\lambda^A,\lambda^B] ,
\ee
where $\omega_{\mu\, B}^{\, \, A}$ is the spin connection, then becomes
the covariant derivative acting on spinors $\pi_\mu=-iD_\mu$. Note
that
\bea
{[}\pi_\mu,\lambda^\nu{]}&=&i\Gamma^\nu_{\mu\rho}\lambda^\rho ,\nn
{[}\pi_\mu,\pi_\nu{]}&=&
-{1\over 2}R_{\mu\nu\rho\sigma}\lambda^\rho\lambda^\sigma .
\eea

The supersymmetry charges take the form
\bea
Q&=&\lambda^\mu(\pi_\mu-G_\mu) ,\nn
Q_{a}&=&\lambda^\mu {J^{(a)\nu}_{\,\, \mu}}(\pi_\nu-G_\nu) .
\eea
Introducing the spin charges
\be
S^a={1\over 2} \lambda^\mu\lambda^\nu {J^{(a)}_{\,\, \mu\nu} } ,
\ee
satisfying
\be
[S^a,S^b]=4\epsilon_{abc}S^c ,
\ee
we have
\bea
Q^a&=&[S^a,Q]\nn
{[}{Q^a},{S^b}{]}&=&\delta^{ab}Q +\epsilon^{abc}Q^c  ,
\eea
The algebra of supercharges is given by
\bea\label{algebra}
\{Q,Q\}&=&2({\cal H}-{\cal Z})  , \nn
\{Q_a,Q_b\}&=&2\,\delta_{ab}({\cal H}-{\cal Z}) , \nn
\{Q,Q_a\}&=&0 .
\eea
where the Hamiltonian ${\cal H}$ and the central charge ${\cal Z}$ is given
by
\bea
&&{\cal H}=
{1\over 2} \biggl( {1\over \sqrt{g}}\pi_\mu \sqrt{g }g^{\mu\nu}\pi_\nu
+ G_\mu G^\mu  + i\lambda^\mu\lambda^\nu D_\mu G_\nu \biggr),\\
&& {\cal Z}= G^\mu \pi_\mu -{i\over 2}  \lambda^\mu\lambda^\nu(D_\mu G_\nu).
\label{hamiltonian}
\eea
Note that the operator $i{\cal Z}$ is  the Lie derivative ${\cal L}_G$
acting on spinors (see {\it e.g.}, \cite{ggpt})
\be
{\cal L}_G\equiv D_G+{1\over 8}[\gamma^\mu,\gamma^\nu]D_\mu G_\nu .
\ee

Although the algebra of supercharges contains a central
charge $\cal Z$ we see that the states will either preserve all
four supersymmetries of the supersymmetric quantum mechanics if 
${\cal H}={\cal Z}$, or none. 
This is entirely consistent with the fact that the parent
$N=2$ field theory has a complex central charge and hence 
BPS states preserve 1/2 of the eight field theory supercharges,
while generic states preserve none of the supersymmetry (of
course the vacuum preserves all of the supersymmetry). 
The BPS bound states satisfying ${\cal H}={\cal Z}$ are 
obtained by finding the normalizable zero modes of the following
Dirac operator on the moduli space,
\begin{equation}
Q{\bf \Psi} =\frac{1}{\sqrt 2}\,\gamma^\mu (-i\nabla_\mu-G_\mu){\bf \Psi}=0 .
\end{equation}

The BPS states of the $N=2$ theory are obtained by solving this 
equation on the monopole
moduli space specified by the Higgs field $b$. The spin content 
of the supermultiplets will be the tensor product of that of 
a half-hypermultiplet,
$(0,0,\pm 1/2)$, which comes from the noninteracting 
center-of-mass fermions, with the spin of the bound states on the relative
moduli space. In the simplest case of a singlet bound state we
get a full hypermultiplet when combined with the corresponding states
from the anti-monopole sector. 

Note that BPS states are only possible if the $\cal Z$ eigenvalue is 
nonnegative, since $\cal H$ is nonnegative\footnote{
This can easily be seen by introducing a complex conjugated operator $Q^*
=\gamma^\mu (+i\nabla_\mu-G_\mu)/\sqrt{2}$. It satisfies the
identity $(Q^*)^2={\cal H}+{\cal Z}$,
so that we have $2{\cal H}=Q^2+(Q^*)^2$,
which is clearly nonnegative.}. From (\ref{geedef}) we see that this 
eigenvalue is given by a linear combination of the electric charges. 
Since classical bosonic dyonic bound states should exist for both
signs of the electric 
charge (recall the BPS bound and the mass bound discussed in section 2), 
we expect ``wrong-sign'' non-BPS dyons as quantum bound states, unless the
potential $G^2/2$ is too weak. 
Such states will solve only the second order Schr\"odinger equation,
\begin{equation}
{\cal H}{\bf \Psi} = {\cal E}{\bf \Psi} , 
\end{equation}
and will break all of the supersymmmetries. Because of this these
states will form longer $N=2$ supermultiplets. For example, 
the smallest possible non-BPS multiplet has degeneracy 16
arising from the 4 states coming from the centre-of-mass fermions with
an additional factor of 4 arising from the supercharges acting 
on the bound states on the relative moduli space. 
This multiplet has highest spin one. It
is identical to the $N=4$ vector multiplet and is a long
multiplet with respect to $N=2$ supersymmetry algebra. 
In the rest of paper, we will consider BPS bound states only.

\section{BPS Dyons in $N=2$ $SU(3)$ Yang-Mills Theory}

We now use the supersymmetric monopole dynamics to analyze the special
case of two distinct monopoles in pure $N=2$ $SU(3)$ Yang-Mills
theory. As we discussed when two Higgs fields are involved
one considers the monopole moduli space determined by $\bf b$ and the
effects of ${\bf a}$ are incorporated via the potential terms.
Recall that for the case of a single Higgs field the
classical $SU(3)$ monopoles can be built 
out of two distinct species of monopoles, known as fundamental monopoles. 
The magnetic charges of these fundamental monopoles correspond
to the two simple roots of $SU(3)$ Lie algebra which are defined 
by the asymptotic behaviour of the Higgs field \cite{ejw}. When two
Higgs fields are involved we use the expectation value
of ${\bf b}$ to specify the simple roots $\balpha$, $\bbeta$
by demanding  ${\bf b}\cdot\balpha\ge 0$ and ${\bf b}\cdot\bbeta\ge 0$.
This is illustrated on the root diagram in figure 1. 
We take the normalization such that $\balpha^2= \bbeta^2=1$ and 
thus $\balpha\cdot\bbeta=-1/2$.

Dyons built on a single $\balpha$ or a single $\bbeta$ magnetic charge 
are easy to find. The moduli space is flat, $R^3\times S^1$, and one
obtains integral electric charges by exciting momentum along the 
internal $U(1)$ angle, which give rise to integral electric charges parallel
to the magnetic charge. 
The possible charges 
$(\hat {\bf g},\hat {\bf q})\equiv ({\bf g}e/4\pi,{\bf q}/e)$ are 
\begin{equation}
(\balpha,n\balpha),\qquad (\bbeta,m\bbeta),
\end{equation}
for integers $n$ and $m$. The potential term in the quantum mechanics is
constant and just contributes to the BPS mass. The quantization of
the free fermions gives a half-hypermultiplet with spin content $(0,0,\pm1/2)$
which combines with the charge conjugate states to form a full
hypermultiplet. 

\vskip 2cm

\begin{center}
\leavevmode
\epsfysize=3in
\epsfbox{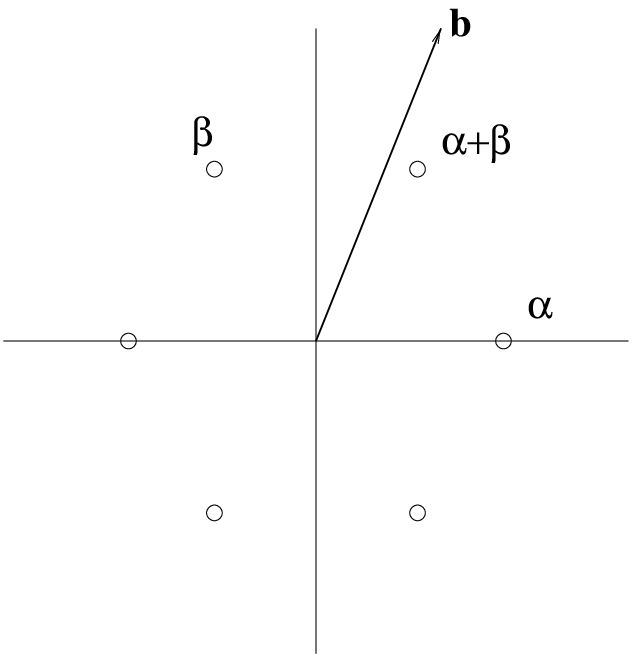}
\end{center}
\begin{quote}
{\bf Figure 1:}  The root diagram of $SU(3)$ Lie algebra. Considering
the Higgs expectation value $\bf b$ as a vector in the root space
defines $\balpha$ and $\bbeta$ to be the simple roots.
\end{quote}
\vskip 1cm

Dyons with magnetic charge $\balpha+\bbeta$ are more difficult to analyze
since the moduli space is now eight-dimensional. The exact metric 
is known \cite{connell,lwy} and it factors into a center of mass piece
and a relative moduli space.
The center of mass part is flat, with metric given by
\begin{equation}
ds^2_{\rm cm} = (\mu_1+\mu_2)\, d{{\vec X}}^2 +
\frac{16\pi^2}{e^4(\mu_1+\mu_2)} d{\xi}^2 ,
\label{cmL}
\end{equation}
where ${\vec X}$ is a three-vector that encodes the center of mass position
of the two monopoles, and $\xi$ is an internal phase. Here we introduced the
masses of the two fundamental monopoles $\mu_1=4\pi {\bf  b}\cdot\balpha/e$
and $\mu_2=4\pi {\bf b}\cdot\bbeta/e$.

The relative moduli space is more complicated and is given 
by the Taub-NUT metric,
\begin{equation}
ds^2_{\rm rel}
= \left(  \mu + \frac{2\pi}{e^2 r}\right ) d{\vec r}^2
+ \frac{4\pi^2/e^4}{\mu + \frac{2\pi}{e^2 r}}
\left (d{\psi} + {\vec w}({\vec r})\cdot d{\vec r}\right)^2 ,
\label{rel}
\end{equation}
where ${\vec r}$ is the relative position vector, while $\psi$ is an angular
coordinate of period $4\pi$. The reduced mass $\mu$ is defined as
$\mu_1\mu_2/(\mu_1+\mu_2)$. The 3-vector ${\vec w}({\vec r}) $ is the Dirac
potential such that $\nabla \times {\vec w}({\vec r}) = -{\vec r}/r^3$.
The eight-dimensional, total moduli space is then given by
\begin{equation}
{\cal M} = R^3 \times \frac{ R^1 \times {\cal M}_{TN}}{\bZ},
\end{equation}
where ${\cal M}_{TN}$ is the Taub-NUT manifold. The identification
map $\bZ$,
\begin{equation}
(\xi, \psi) = (\xi + 2\pi, \psi + \frac{4\pi \mu_2}{\mu_1+\mu_2}),
\end{equation}
arises from the relationships,
\begin{equation}
\xi=\xi_1+\xi_2,\qquad
\psi=2(\mu_1\xi_2-\mu_2\xi_1)/(\mu_1+\mu_2),
\end{equation}
where $\xi_i$ are internal $U(1)$ angles of the two fundamental monopoles
respectively, with each $\xi_i$ having period $2\pi$.

In the special case that $\mu_1=\mu_2$,
$\xi$ becomes periodic by itself with the range of $[0,4\pi)$, and the
moduli space gets simplified a bit, 
\begin{equation}\label{simpabit}
{\cal M} = R^3 \times \frac{ S^1 \times {\cal M}_{TN}}{\bZ_2}.
\end{equation}
The $\bZ_2$ action shifts $\psi$ and $\xi$ by $2\pi$ simultaneously.
The half-integer-quantised momentum along $\xi$ corresponds to overall
$U(1)$ charge in unit of
$e\,(\balpha+\bbeta)$, while the half-integer-quantised
momentum along $\psi$ corresponds to the relative $U(1)$ charge in unit
of $e\,(\balpha-\bbeta)$.
Because of the $\bZ_2$ action, under which the wavefunction
should be invariant, the two momenta are correlated such that either
both are integers or both are half integers.

When the second Higgs expectation $\bf a$ is turned on, the low energy
dynamics is twisted by the tri-holomorphic vector fields.
The moduli space has two tri-holomorphic isometries $K_a$'s, generated by
$\partial /\partial \xi_i$, or equivalently, generated by
$\partial/\partial \psi$ and $\partial/\partial {\xi}$.
The relevant tri-holomorphic vector field $G$ can be decomposed into
two orthogonal pieces,
\begin{equation}
G=e\,{\bf a}\cdot {\bf K}=e\left(\bar a_T\frac{\partial}{\partial \xi}+
 \bar a\frac{\partial}{\partial \psi}\right) ,
\end{equation}
where $\bar a_T$ and $\bar a$ are defined as
${\bf a}\cdot (\balpha+\bbeta)$
and ${\bf a}\cdot (\balpha-\bbeta)$, respectively.
The potential associated with $\partial/\partial {\xi}$ is a constant
and accounts for an important contribution to the total electric (or
excitation) BPS energy when an electric $\balpha+\bbeta$ charge is excited.
The potential associated with
$\partial/\partial \psi$ is position-dependent, and leads to interesting
dyonic bound states. Thus, the interaction between the two fundamental
monopoles is dictated by the low energy supersymmetric quantum mechanics
on Taub-NUT space, twisted by the tri-holomorphic vector field
$\tilde G\equiv e\bar a\, \partial/\partial \psi$.

Given this and the results of the previous section, to determine 
the BPS states with magnetic charge ${\bf g}=4\pi(\balpha+\bbeta)/e$
we need to find normalizable solutions to the Dirac equation
\begin{equation}
\gamma^\mu(-i\nabla_\mu -\tilde G_\mu){\bf \Psi} =0 ,
\end{equation}
where $\nabla$ is the covariant derivative with respect to Taub-NUT space.
This particular problem has been solved by Pope in a
different context \cite{pope}. We can  summarise his results as follows.
Let $\nu$ be the half-integral charge of the state $\bf \Psi$ with respect
to $-i{\cal L}_{\partial_\psi}$. The spectrum is different depending on the
sign of $\bar a\nu$: for negative or vanishing $\bar a\nu$, no normalizable
bound  state exists; for positive $\bar a\nu$, there exists a set of $|2\nu|$
normalizable states, for $|\nu| < (4\pi^2/e^3\mu)|\bar a|$,  
which form an angular momentum $|\nu|-1/2$ multiplet.\footnote{
The restriction of BPS spectrum to $|\nu| < (4\pi^2/e^3\mu)|\bar a|$ 
is reminiscent of 1/4 BPS dyon spectrum in $N=4$ theories 
\cite{bergman,leeyi}.}

One must consider the full moduli space to construct the actual wavefunction.
The electric charges of the state are determined by momenta along $\psi$
and $\xi$. 
Taking into account the $\bZ_2$ identification in 
(\ref{simpabit}) in the 
case that $\mu_1=\mu_2$, the general electric charge is given by
\begin{equation}
{\bf q}=e\,(k+\nu)\,\balpha+e\,(k-\nu)\,\bbeta ,
\end{equation}
where the $\xi$ momentum $k$ is (half-)integer whenever $\nu$ is
(half-)integer. In other words,
\begin{equation}\label{lastone}
{\bf q}=e\,n\,(\balpha+\bbeta)-2e\nu\, \bbeta .
\end{equation}
Although we derived this for the special
case when the masses of the two fundamental monopoles are equal,
this form is valid more generally.

The center of mass part of the moduli space also gives rise
to an additional degeneracy factor. 
The four free fermionic partners to $\vec X$ and $\xi$
generate a half-hypermultiplet structure of degeneracy $4$. Thus,
for each bound state wavefunction of charge $(\nu,k)$, a single BPS
supermultiplet of the highest spin $|\nu|$ is formed. 
The degeneracy of these supermultiplets is summarised below.
\vskip 5mm
\begin{center}
\begin{tabular}{|c|c|c|c|c|c|c|}
\hline
$J_3$ & $\:|\nu|\:$ & $|\nu|-1/2$ & $|\nu|-1$ & $\cdots$
& $-|\nu|+1/2$ & $\:-|\nu|\:$ \\
\hline
Degeneracy & 1 & 2 & 2 & $\cdots$ & 2& 1\\
\hline
\end{tabular}
\end{center}
In particular,
the smallest possible multiplet is the half-hyper multiplet associated
with $\nu=1/2$ (for positive $\bar a$) or $\nu=-1/2$ (for negative $\bar a$).
Combined with the charge conjugate states, it forms a full hypermultiplet
of the $N=2$ theory. We thus find an infinite tower of hypermultiplets
with electric charge given by ${\bf q}=e\,n\,(\balpha+\bbeta)\mp e\, \bbeta$,
for $\bar a$ positive or negative, respectively,
as a subset of the more general BPS states with electric charge 
(\ref{lastone}).

In the limit of $\bar a=0$ it is known that a 
purely magnetic BPS bound state of one $\balpha$ and one
$\bbeta$ monopole does not exist. Indeed such a state 
corresponds to a holomorphic harmonic form on Taub-NUT space and
its existence would have contradicted the $N=4$ S-duality prediction of
a unique anti-self-dual harmonic form. Here we have found this feature 
persists for more general vacua when $\bar a\ne0$.

\section{Consistency with Seiberg-Witten Theory}

In Section 4 we found the complete BPS spectrum of dyons for magnetic
charges $\balpha$, $\bbeta$, and $\balpha+\bbeta$,
for values of $b$ where where the $\balpha$ and $\bbeta$ monopoles are
approximately fundamental. 
These states include dyons in hypermultiplets and we now
comment how this subsector relates to known results about
Seiberg-Witten theory.

The weak coupling spectrum of pure $SU(N)$ theory was studied by
Fraser and Hollowood by analyzing monodromy transformations
on the vacuum moduli space \cite{hollo}. 
Starting with the hypermultiplets
corresponding to fundamental monopoles in regions where only a 
single Higgs field is non-vanishing, the monodromy transformations
predict additional hypermultiplets. For the case of $SU(3)$, 
the vacuum moduli space splits into two disjoint regions separated by a
curve of marginal stability. In one region there are hypermultiplets with
charge vectors $(\hat {\bf g},\hat {\bf q})$ given by
\begin{eqnarray}
&(\balpha, n\balpha), &  \nonumber \\
&(\bbeta,n\bbeta), & \nonumber \\
&(\balpha+\bbeta,n(\balpha+ \bbeta)-\bbeta),
\end{eqnarray}
for any integer $n$. 
In the second region 
the states have identical charges except that the electric charge of the 
third set of states have the form $n(\balpha+ \bbeta)+\bbeta$. 
Consistency then requires that the states with magnetic charge
$\balpha+\bbeta$ decay as they cross the curve of marginal stability. 
This is entirely consistent with 
the spectrum of hypermultiplets that we found in section 4.

The curious asymmetry of spectrum with respect to the sign of electric 
charge also manifests itself in the strong coupling behavior of 
of Seiberg-Witten vacuum moduli space. 
For example, the asymmetry shows
up in the charges of those BPS states that become massless at strong 
coupling singularities. The electromagnetic charges of such states 
in pure $SU(3)$ have been initially worked out explicitly in a symplectic 
basis \cite{lerche}, and later in terms of weight vectors \cite{hollo}.

Let us make some additional observations concerning the singularity
structure of the $SU(3)$ theory. 
Recall that the moduli space  of vacua can be described in terms of spectral
curve of genus two, which can be written in the Weierstrass form as follows
\cite{su3},
\begin{equation}
y^2=\prod_{i=1}^3(x-\phi_i)^2 - \Lambda^6 .
\end{equation}
Here the $\phi_i$'s satisfy the traceless condition $\sum_i \phi_i
=0$ and the vacuum moduli space 
is parameterised by the gauge-invariant combinations
$u\equiv \phi_1\phi_2+\phi_2\phi_3+\phi_3\phi_1$ and
$v\equiv\phi_1\phi_2\phi_3$. 

The $x$ plane has a set of three square-root branch cuts, which, for small
UV cut-off $\Lambda$, are located near the $\phi_i$'s. 
Some of the six end points of
these branch cuts can meet pairwise for special values of the $\phi_i$'s. 
Along such hypersurfaces, a certain charged 
state becomes massless and generates a
singularity of complex co-dimension one on the moduli space. With
$\Lambda\neq 0$, there are six different ways that this can happen in
terms of the $\phi_i$ coordinates, and therefore six different charged states
that can become massless along such hypersurfaces. The six
homology cycles of the Riemann surface 
have some mutual intersection numbers, which is related to the
Schwinger products between the corresponding dyons. From this one can deduce
that the electric and the magnetic charges of these six dyons
are given in a symplectic basis by \cite{lerche},
\begin{eqnarray}
(1,0;1,0)& \quad &(1,0;-1,1) \nonumber \\
(0,1;-1,1)& \quad &(0,1;0,-1) \nonumber \\
(1,1;0,1)& \quad &(1,1;-1,0).
\end{eqnarray}
Here $(m_1,m_2;n_1,n_2)$ represents a charge of the form $ m_1 b_1 + m_2 b_2
+n_1a_1+n_2 a_2$ with $a_i$ and $b_i$ a charge basis whose pairwise
Schwinger products are such that $2 (a_i\odot b_j) =\delta_{ij}$ and
$a_1\odot a_2=0=b_1\odot b_2$. This does not
fix the electromagnetic charges
uniquely. First of all, there are monodromies on the moduli space which
must be fixed by hand. Furthermore, the same intersection matrix shows up
regardless of the number of flavors in the theory, which implies that
there must be certain ambiguity in translating the symplectic basis to
its counterpart in the weight lattice. Part of the ambiguity is resolved 
by noting that near each pair of singularities that extends far out into the
asymptotic region, the local physics should resemble that of $SU(2)$.
Each of the states must thus have a unit magnetic charge that corresponds to
one of three positive roots, $\balpha$, $\bbeta$, $\balpha+\bbeta$.
Finally, when we discuss theories with adjoint matter only, the 
electric charges must also fall onto the root lattice, and this knowledge 
determines the charges up to the monodromies.

Let us choose the monodromy so that $(1,0;1,0)$ is the pure monopole
with charge ($\balpha,0)$ and that $(0,1;0,-1)$ is the pure monopole
with charge ($\bbeta,0)$.
This is possible because
\begin{equation}
(\balpha,0)\odot (\bbeta,0)=\balpha\cdot 0 -0\cdot\bbeta =0 .
\end{equation}
Using the above arguments, 
one then finds the following minimal set of charge vectors on root lattice,
%which are also consistent with the intersection matrix,
\begin{eqnarray}
(\balpha, 0)& \quad &(\balpha,\balpha) \nonumber \\
(\bbeta,-\bbeta)& \quad &(\bbeta,0) \nonumber \\
(\balpha+\bbeta,-\bbeta)& \quad &(\balpha+\bbeta,\balpha).
\end{eqnarray}
In this set of charges, we again find the prominent feature we found in the
low energy dynamics,
namely the asymmetry with respect to the sign of electric charges.
Furthermore, assuming that there is no marginal stability domain wall for
the last two states in passing to asymptotic region we worked in,
this also tells us that there must be dyonic bound states of charges
$(\balpha+\bbeta,-\bbeta)$ and $(\balpha+\bbeta, \balpha)$ in hypermultiplets,
which are exactly the lowest lying bound states found in Section 4.

\section{Conclusions}

We have presented a low energy effective dynamics that allows one to study 
the weak-coupling spectrum of dyons for pure $N=2$ Yang-Mills gauge theory
for general gauge groups in a systematic manner. 
For aligned Higgs fields it has been known for some time that one
should study a supersymmetric quantum mechanics on the BPS monopole moduli
space\cite{gaunt}. For generic vacua, we have argued that the supersymmetric
quantum mechanics is supplemented by a potential term constructed from
the tri-holomorphic Killing vectors on the moduli space. The BPS states
correspond to normalizable zero modes of a Dirac operator twisted
by the Killing vectors. It would be interesting to derive this 
dynamics directly by generalizing the arguments of \cite{gaunt}.

We used the formalism to study the semi-classical BPS spectrum  
for the pure $N=2$ $SU(3)$ gauge theory. The vector ${\bf b}$, defined
by the Higgs vevs and by the choice of the 
Weyl chamber, %the electric and magnetic charge vectors, 
specifies a basis of simple roots in the algebra. 
When the magnetic charge is given by a simple root, $\balpha$ or $\bbeta$,
there is a tower of dyons in hypermultiplets with parallel electric
charges. A more interesting structure emerges for magnetic charge
given by $\balpha+\bbeta$. Firstly, the electric charge vector is 
necessarily not parallel to the magnetic charge and in particular, 
there is no purely magnetic BPS state of this charge. 
Secondly, there is a tower of hypermultiplets with electric charge that are
consistent with previous results in Seiberg-Witten theory in that they
are in accord with semi-classical monodromy and strong coupling singularities.
More generally, the hypermultiplets are a special case 
of an infinite tower of BPS states with 
electric charge $(m,l)$, with either $m>l$ or $m<l$, depending
on the sign of the second Higgs ${\bf a}$, with maximal spin $|m-l|/2$.

It would be interesting to extend these results for more general magnetic
charge, but due to the lack of understanding of the relevant
$SU(3)$ monopole moduli spaces this appears difficult. 
A more accessible problem would be to generalise the results
of this paper to analyse the BPS spectrum for $SU(N)$ gauge group
with magnetic charge given by $(1,1,\dots,1)$.

The effective Lagrangian in this paper is for the case
when hypermultiplets in the $N=2$ Yang-Mills theory can be ignored.
If there are light hypermultiplets, one must introduce new quantum mechanical
degrees of freedom associated with them. Presumably, there is a way to couple
them to our Lagrangian while preserving the four real supercharges. However,
it goes beyond the scope of this paper, and will be worked out elsewhere.

%%%%%%%%%%%%%%%%%%%%%%%%%%%%%%%%%%%%%%%%%%%%
\medskip
\section*{Acknowledgments}
\noindent
We would like to thank Bobby Acharya, Kimyeong Lee, and David Tong for 
discussions. JPG is supported in part by an EPSRC Advanced Fellowship,
NK is supported on a joint EPSRC/PPARC PDRA and both are
supported in part by PPARC through SPG \#613. NK would also like
to thank KIAS for hospitality while this work was in progress.
JP is supported in part by U.S. Department of 
Energy Grant DOE DE-FG02-90-ER40542. 

%%%%%%%%%%%%%%%%%%%%


\medskip


\begin{thebibliography}{99}

\bibitem{gaunt}
J.P.~Gauntlett,
%``Low-energy dynamics of N=2 supersymmetric monopoles,''
Nucl.\ Phys.\ {\bf B411} (1994) 443
hep-th/9305068.

\bibitem{blum}
 J. Blum, Phys. Lett. {\bf B333} (1994) 92, hep-th/9401133.

\bibitem{sen}
A. Sen, Phys.Lett.{\bf  B329} (1994) 217, hep-th/9402032.

\bibitem{sethi}
S. Sethi, M. Stern and E. Zaslow, Nucl. Phys. {\bf B457} (1995) 484,
hep-th/9508117; J. P. Gauntlett and J. A. Harvey,
Nucl. Phys. {\bf B463} (1996) 287, hep-th/9508156.

\bibitem{lwy}

J.P. Gauntlett and D.A. Lowe, Nucl. Phys. {\bf B472} (1996) 194;
K. Lee, E.J. Weinberg, and P. Yi, Phys. Lett. {\bf B376} (1996) 97;
Phys. Rev. {\bf D54} (1996) 1633.


\bibitem{gibbons}
G. W. Gibbons, Phys. Lett. {\bf B382} (1996) 53, hep-th/9603176.


\bibitem{hf}
C.~Fraser and T.J.~Hollowod,
Phys.\ Lett.\ {\bf B402} (1997) 106,
hep-th/9704011.

\bibitem{bergman} O. Bergman, 
Nucl. Phys. {\bf B525} (1998) 104, hep-th/9712211;    O. Bergman and B. Kol,
Nucl. Phys. {\bf B536} (1998) 149, hep-th/9804160.

\bibitem{hashimoto}
K. Hashimoto, H. Hata, and N. Sasakura,
Phys. Lett. {\bf B431} (1998) 303, hep-th/9803127;
T. Kawano and K. Okuyama, Phys. Rev. {\bf D60} (1999) 
046005, hep-th/9901107; K. Hashimoto, H. Hata, and N. Sasakura,
 Nucl.Phys. {\bf B535} (1998) 83, hep-th/9804164.

\bibitem{leeyi}
K.~Lee and P.~Yi,
%``Dyons in N = 4 supersymmetric theories and three-pronged strings,''
Phys.\ Rev.\ {\bf D58} (1998) 066005
hep-th/9804174.

\bibitem{bak}
D. Bak, K.  Hashimoto, B. Lee, H. Min, and N. Sasakura, 
Phys.Rev. {\bf D60} (1999) 046005, hep-th/9901107.


\bibitem{tong}
D.~Tong,
%``A note on 1/4-BPS states,''
Phys.\ Lett.\ {\bf B460} (1999) 295,
hep-th/9902005.

\bibitem{blly}
D.~Bak, C.~Lee, K.~Lee and P.~Yi,
{\it Low energy dynamics for 1/4 BPS dyons,}
hep-th/9906119; D.~Bak, K.~Lee and P.~Yi, {\it Quantum 1/4 BPS dyons,}
hep-th/9907090.
D. Bak and K. Lee, {\it 
Comments on the Moduli Dynamics of 1/4 BPS Dyons},
hep-th/9909035.

 



\bibitem{connell} {S.A. Connell, {\it The Dynamics of SU(3) Charge
(1,1) Magnetic Monopoles}, University of South Australia Preprint
(1994).}

\bibitem{wittenolive}
E.~Witten and D.~Olive,
%``Supersymmetry Algebras That Include Topological Charges,''
Phys.\ Lett.\ {\bf 78B} (1978) 97.

\bibitem{pope}
C.N. Pope, Nucl. Phys. {\bf B141} (1978) 432. 

\bibitem{sw}
N. Seiberg and  E. Witten, Nucl. Phys. {\bf B426} (1994) 19; Erratum-ibid. 
{\bf B430} (1994) 485, hep-th/9407087.


\bibitem{hollo}
C.~Fraser and T.J.~Hollowood,, Nucl.Phys. {\bf B490} (1997) 217,
hep-th/9610142; T.J. Hollowood, Nucl.Phys. {\bf B517} 
(1998) 161, hep-th/9705041.


\bibitem{gkmtz}
J.P. Gauntlett, C. Koehl, D. Mateos, P. Townsend and M. Zamaklar,
%Finite energy Dirac-Born-Infeld monopoles and string junctions,
Phys.Rev. {\bf D60} (1999) 045004, hep-th/9811024.

\bibitem{ejw} E. Weinberg,  Nucl. Phys. {\bf B167} (1980) 500.
%U(1)^n and fundamental monopoles



\bibitem
{ggpt}
J.P. Gauntlett, G.W. Gibbons, G. Papadopoulos and P.K. Townsend,
Nucl.Phys. {\bf B500} (1997) 133, hep-th/9702202.

\bibitem{su3} 
A. Klemm, W. Lerche, S. Theisen and S. Yankielowicz,
Phys.Lett. {\bf B344} (1995) 169, hep-th/9411048;
Philip C. Argyres and Alon E. Faraggi, Phys. Rev. Lett. 
{\bf 74} (1995) 3931. 


\bibitem{lerche}
A. Klemm, W. Lerche, S. Theisen and S. Yankielowicz,
{\it On the Monodromies of N=2 Supersymmetric Yang-Mills Theory},
hep-th/9412158.



\end{thebibliography}
\end{document}